\documentclass[preprint,12pt]{elsarticle}

\usepackage{amssymb}
\usepackage{amsthm}
\usepackage{amsmath}
\usepackage{chngpage}

\journal{arxiv}

\begin{document}

\begin{frontmatter}



\title{Empirical Evidence of Isospin Memory in Compound Nuclear Fission}


\author[1]{Swati Garg\corref{mycorrespondingauthor}}
\ead{swat90.garg@gmail.com}
\author[1,2]{and Ashok Kumar Jain}
\address[1]{Department of Physics, Indian Institute of Technology Roorkee, Roorkee-247667, India}
\address[2]{Amity Institute of Nuclear Science \& Technology, Amity University Uttar Pradesh, Noida, India.}
\begin{abstract}
We present empirical evidence of isospin dependence in the compound nuclear fission cross-sections and fission widths, which suggests that the compound nucleus (CN) possibly retains the memory of the isospin when it is formed. We examine the idea, first proposed by Yadrovsky~\cite{yadrovsky}, for three pairs of reactions where experimental data of fission cross section at various excitation energies are available. One of the pairs of reactions is the same as used by Yadrovsky i.e.  $^{209}$Bi($p$, f) and $^{206}$Pb($\alpha$, f) leading to the CN $^{210}$Po but with an improved experimental data set. The other two pairs of reaction sets are, $^{185}$Re($p$, f) and $^{182}$W($\alpha$, f) leading to the CN $^{186}$Os and, $^{205}$Tl($p$, f) and $^{202}$Hg($\alpha$, f) leading to the CN $^{206}$Pb. An observable difference between the fission branching ratios in two different isospin states suggests that the CN seems to remember its isospin at the point of formation. This possibility is further supported by another method, where additional empirical evidence for four CN, viz. $^{210}$Po, $^{209}$Bi, $^{207}$Bi, and $^{198}$Hg, is obtained from the experimental data in Zhukova et al.~\cite{ignatyuk1977b}. Further, the data also suggest a possible new signature of the weakening of CN process and gradual transition to non-compound processes as the energy rises. Fresh experimental efforts as proposed, are required to confirm these findings.

\end{abstract}


\end{frontmatter}


\section{Introduction}
\label{sec 1}
Since the discovery of isospin by Heisenberg in 1932~\cite{heisenberg}, isospin has emerged as a useful quantum number in both nuclear and particle physics. Isospin was included as a third coordinate in nucleonic wave functions along with spin and orbital angular momentum coordinates a long time ago~\cite{cassen}. Till 1960's, isospin was assumed to be useful only in the light mass nuclei as isospin purity decreases when we move towards heavier nuclei. However, in 1962, Lane and Soper~\cite{lane} showed theoretically that isospin may remain a good quantum number in heavy nuclei which are naturally richer in neutrons. Later, many theoretical calculations have been reported using various methods to show that as we move towards heavy mass nuclei which have more neutrons than protons ($N>Z$), isospin mixing of states starts decreasing leading to more pure isospin states~\cite{sliv,bohr,auerbach}. Very recently, Loc, Auerbach and Colo~\cite{loc} have presented a detailed study of isospin mixing in ground states of even-even nuclei, making a distinction between Coulomb mixing and isospin mixing. They conclude that isospin mixing remains very small while Coulomb mixing may become large in heavy nuclei. The neutron excess in heavy mass nuclei leads to large $T$ values, which reduces the isospin mixing by a factor of $1/(T+1)$, as also concluded by Lane and Soper~\cite{lane}. 

We have recently applied the idea of isospin purity in heavy mass nuclei which are naturally neutron-rich systems ($N>Z$) to reproduce the fission fragment distributions~\cite{jain,swati,jain1,swati1,swati3}. We have calculated the relative yields of fission fragments in heavy ion and thermal neutron induced reactions based on the premise of conservation of isospin in CN fission with remarkable success. These results confirm the validity of isospin in $N>Z$ nuclei and constitute a direct evidence of goodness of isospin in neutron-rich systems.

In this work, we present empirical evidence which further supports the significant role being played by isospin in the heavy mass region, and is based on the idea proposed by Yadrovsky~\cite{yadrovsky}. Yadrovsky, in 1975, considered a set of proton and alpha induced reactions, namely $^{209}$Bi($p$, f) and $^{206}$Pb($\alpha$, f) leading to the same CN $^{210}$Po, but populated in different isospin states.  The fission decay widths from two different isospin states of CN at various excitation energies were then compared~\cite{yadrovsky}. The difference between the fission decay widths in different isospin states was found to be quite large,  and it was concluded that ``a nucleus remembers the isospin values of the nuclear states leading to fission".

In this paper, we revisit the idea of Yadrovsky~\cite{yadrovsky} with more precise and newer data sets. We calculate the fission branching ratios from two isospin states of CN using an approach very similar to that used by Yadrovsky for three pairs of reactions; one pair of reactions is the same as used by Yadrovsky but with an improved data set. The other two pairs are $^{185}$Re($p$, f) and $^{182}$W($\alpha$, f), and $^{205}$Tl($p$, f) and $^{202}$Hg($\alpha$, f). The data of fission cross-sections for the CN $^{210}$Po and $^{186}$Os were taken from Ignatyuk et al.~\cite{ignatyuk}. However, the data of fission cross-sections for the CN $^{206}$Pb were taken from Ignatyuk et al.~\cite{ignatyuk} and Moretto et al.~\cite{moretto}. Using the CN formation cross-sections, $\sigma_{p}$ and $\sigma_{\alpha}$ from the PACE4 code~\cite{tarasov}, we obtained the fission branching ratios for the two isospin states of the same CN. From the results of all three calculations, it is found that a strong isospin dependence is evident from the large differences in the fission decay widths, an important structural property of nuclei.

We note that an anomalous behavior, similar to that discussed above, was observed by Ignatyuk et al.~\cite{ignatyuk1977b,ignatyuk} in seven more cases. However, three of the seven cases have data only at lower energies. We, have, therefore analyzed the experimental data from Zhukova et al.~\cite{ignatyuk1977b} for the remaining four pairs of reactions, $^{209}$Bi($p$, f) and $^{206}$Pb($\alpha$, f) forming CN $^{210}$Po; $^{208}$Pb($p$, f) and $^{205}$Tl($\alpha$, f) forming CN $^{209}$Bi; $^{206}$Pb($p$, f) and $^{203}$Tl($\alpha$, f) forming CN $^{207}$Bi; $^{197}$Au($p$, f) and $^{194}$Pt($\alpha$, f) forming CN $^{198}$Hg. These data sets were directly analyzed to obtain empirical fission branching ratios in the two isospin states. These results also suggest an isospin dependence as claimed in the previous three cases.  

Since our calculations assume a compound nucleus formation and its decay, the results are valid only up to the excitation energies where fission proceeds mainly via the compound nuclear processes. At higher energies, the non-compound processes start making contribution to the total cross-section and the contribution of CN fission begins to decline leading to change in the values of fission widths. These non-compound processes mainly include pre-equilibrium and other modes of fission with increasing energy~\cite{koning}. Other non-compound processes like scattering and particle emission are always present. Therefore, we also correlate the observed changes in the fission width with the decline of the compound nuclear processes in nuclear fission. A preliminary report of the initial work has been presented by us in Ref.~\cite{swati2}.

\section{Isospin dependence of fission branching ratios}
\label{sec 2}
We consider two types of reactions, one with the projectile having $T_3=0$ (as in alpha particle) and the other having $T_3= -1/2$ (as in proton) but both leading to the same CN, albeit in different isospin states. It is possible to have many such combinations of projectiles such as proton and alpha, proton and deuteron, $^3$He and alpha etc., which may lead to the formation of the same CN but in different isospin states. Surprisingly, there are not many data sets of this kind in the literature. The lack of target-projectile combinations, which are practically feasible, have probably limited the availability of such complete measurements. In this paper, we have chosen the combination of proton and alpha induced fission, where several data sets already exist. In the case of proton induced fission, a proton with energy $E_p$ and isospin $T=1/2$, $T_3=-1/2$ is incident on a target having $T=T_3=T_0$. From isospin conservation, there are two possible isospin states of CN allowed by the isospin algebra. Thus, the CN is formed in two isospin states $T_{CN}=T_0+1/2$ and $T_0-1/2$ with cross-sections $\sigma_{p}^{T_0+1/2}$ and $\sigma_{p}^{T_0-1/2}$, respectively, and with same $T_{3_{CN}}=T_0-1/2$. The average fission cross section may be written as a sum of two terms corresponding to these two isospin states,
\begin{equation}
\langle \sigma_{p,f} \rangle = {\sigma_{p}^{T_0+1/2}} {\dfrac{\Gamma_{f}^{T_0+1/2}}{\Gamma_{total}^{T_0+1/2}}}+{\sigma_{p}^{T_0-1/2}} {\dfrac{\Gamma_{f}^{T_0-1/2}}{\Gamma_{total}^{T_0-1/2}}}
\end{equation}
where $\Gamma_{f}^{T_0+1/2}$ and $\Gamma_{f}^{T_0-1/2}$ are the fission widths from $T_0+1/2$ and $T_0-1/2$ states of CN, respectively, whereas, $\Gamma_{total}^{T_0+1/2}$ and $\Gamma_{total}^{T_0-1/2}$ are the corresponding total decay widths from these two states. Similarly, in the case of alpha induced fission, when an $\alpha$-particle with energy $E_{\alpha}$ is incident on a target with isospin $T'=T_3'=T_0-1/2$, the CN is formed in only one possible isospin state, $T_{CN}= T_0-1/2$ with cross-section $\sigma_{\alpha}^{T_0-1/2}$. Therefore, the average fission cross section has only one term,
\begin{equation}
\langle \sigma_{\alpha,f} \rangle = {\sigma_{\alpha}^{T_0-1/2}} {\dfrac{\Gamma_{f}^{T_0-1/2}}{\Gamma_{total}^{T_0-1/2}}}
\end{equation}
where $\Gamma_{f}^{T_0-1/2}$ and $\Gamma_{total}^{T_0-1/2}$ are the fission width and total decay width from CN, respectively. As a result, the following expressions of fission branching ratios for $T_0+1/2$ and $T_0-1/2$ states of CN are easily obtained,
\begin{equation}
\dfrac{\Gamma_{f}^{T_0-1/2}}{\Gamma_{total}^{T_0-1/2}}=\dfrac{\langle \sigma_{\alpha,f} \rangle} {\sigma_{\alpha}^{T_0-1/2}}=\dfrac{\langle \sigma_{\alpha,f} \rangle} {\sigma_{\alpha}}
\end{equation}
and,
\begin{equation}
\dfrac{\Gamma_{f}^{T_0+1/2}}{\Gamma_{total}^{T_0+1/2}}=(2T_0+1) \bigg[\dfrac{\langle \sigma_{p,f} \rangle} {\sigma_{p}}-\dfrac{\langle \sigma_{\alpha,f} \rangle} {\sigma_{\alpha}}\bigg]
\end{equation}
where $\sigma_{p}$ and $\sigma_{\alpha}$ are the formation cross-sections of CN in proton and alpha induced reactions, respectively. Yadrovsky calculated the fission branching ratios by using the ratios $\dfrac{\langle \sigma_{p,f} \rangle} {\sigma_{p}}$ and $\dfrac{\langle \sigma_{\alpha,f} \rangle} {\sigma_{\alpha}}$ from Gadioli $\it{et}$ $\it{al.}$~\cite{gadioli} and showed that there is a difference of the order of 100 between the fission branching ratios from two isospin states of the CN. He, further, calculated the individual fission decay widths $\Gamma_{f}^{T_0+1/2}$ and $\Gamma_{f}^{T_0-1/2}$ and found a difference of the order of $10^4$-$10^5$ between the two, which lead him to conclude that ``a nucleus remembers the isospin values of the nuclear states leading to fission"~\cite{yadrovsky}. However, in one of the approaches, we calculate the fission branching ratios by using the values of $\langle \sigma_{p,f} \rangle$ and $\langle \sigma_{\alpha,f} \rangle$ in Eq. (3) and (4) directly from the experimental data.

We calculate $\sigma_{p}$ and $\sigma_{\alpha}$ at different excitation energies by using the statistical code PACE4~\cite{tarasov}. It basically performs statistical equilibrium model calculations. The CN formation cross-section by fusion are calculated in this code by using the model of Bass~\cite{bass}, which generally provides good estimates of the experimental data. Also, it can calculate the fusion cross-section below the Coulomb barrier using the quantum mechanical approach~\cite{wong}. In this code, the transmission coefficients for light particle evaporation ($n$, $p$, $\alpha$) are obtained using optical model calculations~\cite{perey,huizenga}. The level density in PACE4 is determined by three factors; (i) the $a$ parameter which is the ratio of number of nucleons ($A$) and a constant factor $K$ (ii) the ratio of $a$ at saddle point and ground state deformations, $a_f/a$ (iii) the fission barrier $B_f$ which is the product of a constant factor and the rotating liquid drop fission barrier. We note that the $\sigma_{p}$ values remain nearly constant and $\sigma_{\alpha}$ exhibits a very slow variation in the energy region being considered. Both the cross-sections remain unaffected by small changes in the input parameters like the level density, number of events, and the ratio of level density at saddle to ground state. The results obtained from PACE4 may be directly compared with the experimental cross-sections. However, this code is not suitable for the pre-equilibrium and incomplete fusion processes, which begin to dominate at higher energies. This is the region where we find a significant shift in the nature of fission widths, providing us a possible signature of the dominance of non-compound processes at higher energies.

\section{Experimental Data}
\label{sec:exp data}
\begin{figure}
\begin{center}
\begin{adjustwidth}{-0.6in}{-0.6in}
\includegraphics[width=13cm,height=10cm]{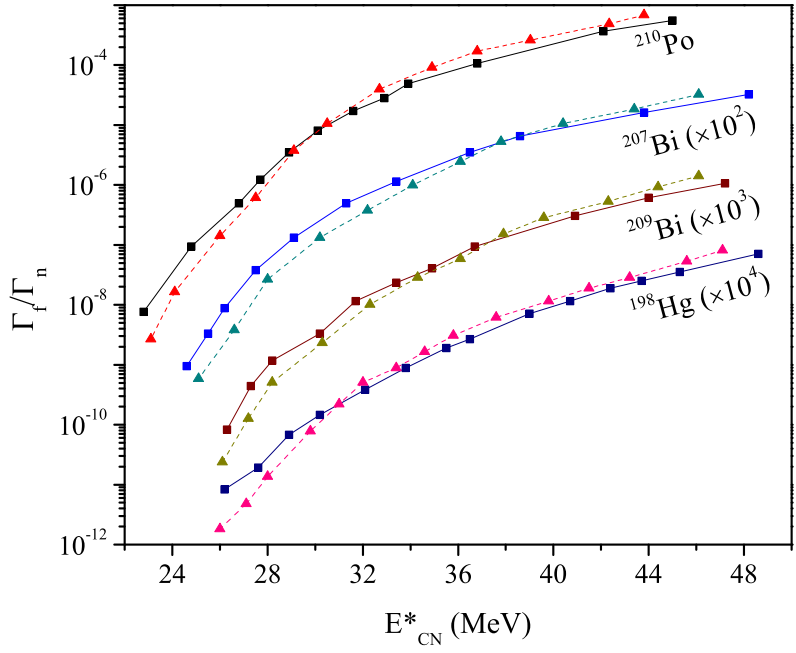}
\caption{\label{fig:igna1977}(Color online) Comparison of the ratio of fission decay width to neutron decay width, $\dfrac{\Gamma_{p,f}}{\Gamma_n}$ (solid line) and $\dfrac{\Gamma_{\alpha,f}}{\Gamma_n}$ (dashed line) for the CN $^{210}$Po, $^{207}$Bi, $^{209}$Bi and $^{198}$Hg. The figure is adapted from Zhukova $\it{et}$ $\it{al.}$~\cite{ignatyuk1977b}.}
\end{adjustwidth}
\end{center}
\end{figure}

	~~~~~Ignatyuk $\it{et}$ $\it{al.}$ have reported a number of measurements for proton and alpha induced fission cross-sections in a series of papers~\cite{ignatyuk1977b,ignatyuk,ignatyuk1976,ignatyuk1977a}. The isochronous cyclotron of the Nuclear Physics Institute of the Kazakh Academy of Sciences was used to measure the fission cross-sections. In the first paper~\cite{ignatyuk1976}, the measurements of the cross-sections for ($\alpha$, f) were made for 29 target nuclei from Hf to Bi. In their second paper~\cite{ignatyuk1977a}, the angular distribution of fragments emitted in ($\alpha$, f) reactions have been reported. In the third paper~\cite{ignatyuk1977b}, Zhukova $\it{et}$ $\it{al.}$ have reported the measurements of cross-sections for ($p$, f) reactions of 15 target nuclei from W to Bi. Further analysis was done to determine the fission barriers and the dependence of fission probability on angular momentum. In this paper, they also compare the fissility for ($\alpha$, f) and ($p$, f) reactions. The fissility is defined as the ratio of fission cross-section to CN formation cross-section, which is nearly the same as the ratio of the respective widths,
\begin{equation}
\dfrac{\sigma_f}{\sigma_{CN}} \approx \langle \dfrac{\Gamma_f}{\Gamma_n} \rangle 
\end{equation}
where $\Gamma_f$ and $\Gamma_n$ are fission and neutron decay widths, respectively. We have presented a comparison of the fissility of ($\alpha$, f) and ($p$, f) reactions for four CN in Fig.~\ref{fig:igna1977} \cite{ignatyuk1977b}. An anomalous behavior in the fissility of the CN $^{210}$Po, $^{209}$Bi, $^{207}$Bi, $^{198}$Hg is observed for all the four cases, which is similar to the effect discussed by us. At lower energies, the fissility of ($p$, f) is greater than ($\alpha$, f). At higher energies, the inverse inequality occurs and the fissility of ($p$, f) becomes smaller than ($\alpha$, f), which is in accordance with the normally expected angular momentum dependence of fission. A comparison for three more cases having the CN $^{186}$Os, $^{189}$Ir and $^{191}$Ir is presented by Zhukova $\it{et}$ $\it{al.}$~\cite{ignatyuk1977b}. However, in these three cases, the data are available only at lower energies.

Ignatyuk $\it{et}$ $\it{al.}$  have discussed this anomalous behavior in their fourth paper in a more detailed manner for the cases of the CN $^{210}$Po and $^{186}$Os, where data are most complete and available over a wider range of energy [19]. Herein, they have made corrections for the non-compound processes, angular momentum distribution, and emission of neutrons before fission. Even after these corrections, the anomalous behavior in favor of ($p$, f) reactions persists. This behavior changes sign at higher energy and the difference between ($\alpha$, f) and ($p$, f) becomes large at higher excitation energies. Assuming an angular momentum distribution on excitation energy, they could bring ($p$, f) closer to ($\alpha$, f) at lower energies but not completely. The ratio of alpha to proton fissility was found to be about 0.5 at lower energies, which became 1 around 28-30 MeV and more than 1 at higher energies. The same angular momentum distribution could not explain the behavior at higher energies and they had to assume a reduced angular momentum distribution at higher energies. 

In the next section, we present an analysis of these data using two distinct methods. In the first method, we use the experimental data on fission cross-sections $\sigma_{p,f}$ and $\sigma_{\alpha,f}$ for three pairs of reactions. The first two pairs correspond to the CN $^{210}$Po and $^{186}$Os. The data for these two cases have been taken from Ignatyuk $\it{et}$ $\it{al.}$~\cite{ignatyuk}. The third pair of reactions leads to the formation of CN $^{206}$Pb. We have taken the data of fission cross-section for $^{202}$Hg($\alpha$, f) from Moretto $\it{et}$ $\it{al.}$~\cite{moretto} and for $^{205}$Tl($p$, f) from Ignatyuk $\it{et}$ $\it{al.}$~\cite{ignatyuk}. Both lead to the CN $^{206}$Pb. In Moretto $\it{et}$ $\it{al.}$~\cite{moretto}, fission excitation functions are reported for 14 CN from $A=186$ to $A=213$. Since, we are taking the data from two different sources, we must normalize the yields for comparison. Prokofiev has pointed out (see comment A2 of Ref.~\cite{prokofiev}) that the dataset from Ignatyuk $\it{et}$ $\it{al.}$~\cite{ignatyuk} needs to be re-normalized with a factor of 0.449$\pm$0.041 so that it may be compared with the data from Moretto $\it{et}$ $\it{al.}$~\cite{moretto}. We, therefore, re-normalize the data of 2$^{205}$Tl($p$, f) from Ignatyuk $\it{et}$ $\it{al.}$~\cite{ignatyuk} by using factors 0.4 (the lower limit) and 0.5 (the upper limit).

In analyzing the data sets for the three CN, viz. $^{210}$Po, $^{186}$Os and $^{206}$Pb, we need the cross-sections $\sigma_{p}$ and $\sigma_{\alpha}$, which are calculated from PACE4~\cite{tarasov}. We may, therefore, term this method as empirical approach 1.

In the second method, we directly use the data presented in Fig.~\ref{fig:igna1977} for the four CN viz. $^{210}$Po, $^{209}$Bi, $^{207}$Bi and $^{198}$Hg. In this method, we do not require the PACE4 calculations. Therefore, the results are purely empirical in nature and we term it as empirical approach 2. A detailed discussion of these two approaches and the results are presented in the next section.
\section{Results and Discussion}
\label{sec 3}
\textbf{Empirical approach 1 -} We have analyzed three pairs of reaction data sets. First, we consider the same set of reactions as used by Yadrovsky, i.e. $^{209}$Bi($p$, f) and $^{206}$Pb($\alpha$, f). Both lead to the formation of the same CN, $^{210}$Po. Yadrovsky had taken the experimental data for both the reactions from Gadioli $\it{et}$ $\it{al.}$~\cite{gadioli}. However, Prokofiev~\cite{prokofiev} has pointed out that this data set is not reliable due to normalization issues. Further, Gadioli $\it{et}$ $\it{al.}$~\cite{gadioli} report only the ratios $\sigma_{p,f}/\sigma_{p}$ and $\sigma_{\alpha,f}/\sigma_{\alpha}$ whereas Ignatyuk $\it{et}$ $\it{al.}$~\cite{ignatyuk} report the cross-sections $\sigma_{p,f}$ and $\sigma_{\alpha,f}$ explicitly. Therefore, we have taken the experimental data of fission cross-sections at various excitation energies of CN from Ignatyuk $\it{et}$ $\it{al.}$~\cite{ignatyuk} for both the reactions. It should be noted that in some papers, the fission cross-section data are relative while in others, these are absolute. Therefore, we have used the experimental data for both the reactions from the same paper~\cite{ignatyuk} so that there are no issues related to the normalization of data. The extracted data for $\sigma_{p,f}$ and $\sigma_{\alpha,f}$ are shown in Table 1 at four different energies. We then calculate $\sigma_{p}$ and $\sigma_{\alpha}$ for both the reactions by using the code PACE4. As it may be seen, the values of $\sigma_{p}$ remain nearly constant while $\sigma_{\alpha}$ changes very little over the energy range considered.

\begin{table*}
\centering
\caption{The table lists the incident energies of proton ($E_p$) and alpha ($E_{\alpha}$), excitation energies of CN ($E^{*}_{CN}$), fission cross-section data $\sigma_{p,f}$ and $\sigma_{\alpha,f}$ extracted from Ignatyuk $\it{et}$ $\it{al.}$~\cite{ignatyuk} for a pair of $^{209}$Bi($p$, f) and $^{206}$Pb($\alpha$, f) reactions, CN formation cross-section $\sigma_{p}$ and $\sigma_{\alpha}$ and fission branching ratios $\dfrac{\Gamma_{f}^{T_0+1/2}}{\Gamma_{total}^{T_0+1/2}}$ and $\dfrac{\Gamma_{f}^{T_0-1/2}}{\Gamma_{total}^{T_0-1/2}}$ for $T_0+1/2$ and $T_0-1/2$ states of CN.}
\label{tab-1}       
\begin{tabular}{|l|l|l|l|l|l|l|l|l|}
\hline
\rule{0pt}{20pt} $E_{p}$ & $E_{\alpha}$ & $E^{*}_{CN}$ &  $\sigma_{p,f}$ & $\sigma_{\alpha,f}$ & $\sigma_{p}$ & $\sigma_{\alpha}$ & $\dfrac{\Gamma_{f}^{T_0+1/2}}{\Gamma_{total}^{T_0+1/2}}$ & $\dfrac{\Gamma_{f}^{T_0-1/2}}{\Gamma_{total}^{T_0-1/2}}$ \\(MeV) & (MeV) & (MeV) & ($\mu$b ) & ($\mu$b) & (mb) & (mb) &&\\
\hline
\rule{0pt}{10pt} 22.6 & 33.5 & 27.5 & 15.4 & 12.5 &  $1.39 \times 10^{3}$ & $1.29 \times 10^{3}$ & $6.14 \times 10^{-5}$ & $9.67 \times 10^{-6}$\\
\hline
\rule{0pt}{10pt} 23.7 & 34.6 & 28.6 & 52.6 & 38.2 & $1.41 \times 10^{3}$ & $1.35 \times 10^{3}$ & $3.96 \times 10^{-4}$ & $2.83 \times 10^{-5}$\\
\hline
\rule{0pt}{10pt} 24.6 & 35.6 & 29.5 & 93.9 & 96.8 & $1.42 \times 10^{3}$ & $1.39 \times 10^{3}$ & $-1.53 \times 10^{-4}$ & $6.96 \times 10^{-5}$\\
\hline
\rule{0pt}{10pt} 27.5 & 38.6 & 32.4 & 474.4 & 585.2 & $1.44 \times 10^{3}$ & $1.52 \times 10^{3}$ & $-2.45 \times 10^{-3}$ & $3.85 \times 10^{-4}$\\
\hline
\end{tabular}
\end{table*}

\begin{figure}
\begin{adjustwidth}{-0.6in}{-0.6in}
\includegraphics[width=17cm,height=14cm]{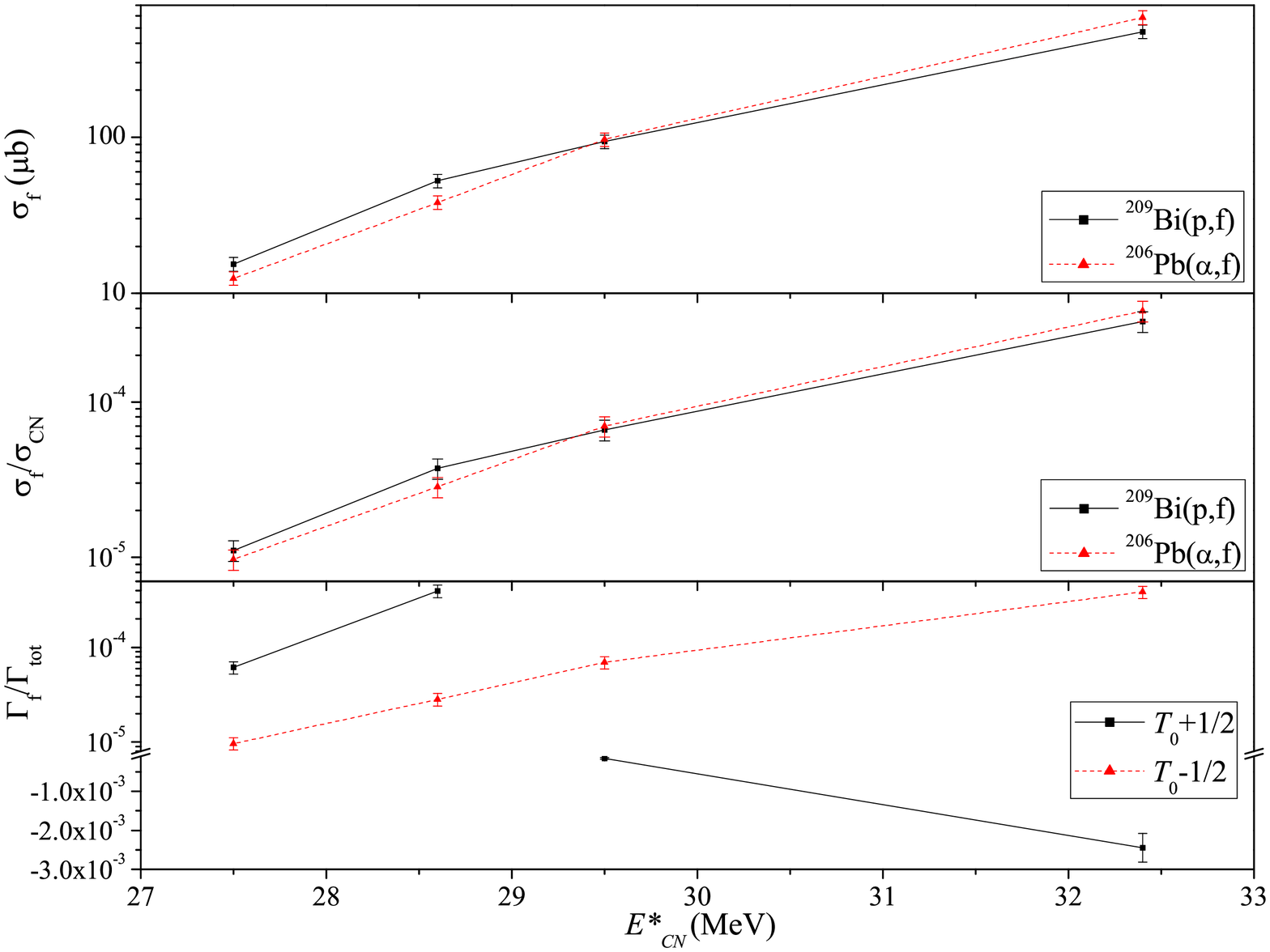}
\caption{\label{fig:210po}(Color online) Comparison of fission cross-section, $\sigma_{p,f}$ and $\sigma_{\alpha,f}$ (upper panel), ratio of fission cross-section to total cross-section, $\dfrac{\sigma_{p,f}} {\sigma_{p}}$ and $\dfrac{\sigma_{\alpha,f}} {\sigma_{\alpha}}$ (middle panel), and the ratio of fission width to total decay width from $T_0+1/2$ and $T_0-1/2$ isospin states of CN, $\dfrac{\Gamma_{f}^{T_0+1/2}}{\Gamma_{total}^{T_0+1/2}}$ and $\dfrac{\Gamma_{f}^{T_0-1/2}}{\Gamma_{total}^{T_0-1/2}}$ (lower panel) for a pair of $^{209}$Bi($p$, f) and $^{206}$Pb($\alpha$, f) reactions, both leading to the same compound nucleus $^{210}$Po. In the lower panel, we have two scales, linear scale before break and log scale after break. We note that the uncertainties in the data points in the topmost panel are of the order of 10\% and in the middle panel are of the order of 15\%.}
\end{adjustwidth}
\end{figure}

We plot the measured cross-sections for $\sigma_{p,f}$ and $\sigma_{\alpha,f}$ in Fig.~\ref{fig:210po} (upper panel). In the middle panel, we plot the ratio of the fission cross-sections to the total cross-section. In the lower panel, we plot the ratio of fission decay width to total decay width for two isospin states of CN, $T_0+1/2$ and $T_0-1/2$, obtained by using the Eqs. (3) and (4). We notice a difference on the order of 10 between the fission branching ratios from the two isospin states which is a clear indication of isospin dependence in fission. The experimental data on fission cross-sections generally carries an error of about 5\%. We raise this error to 10\% as suggested by the evaluators and accordingly plotted in the upper panel of Fig.~\ref{fig:210po}. We also assume a 10\% error in the CN formation cross-sections $\sigma_{p}$ and $\sigma_{\alpha}$ calculated by PACE4. We take an error of about 15\% in the middle panel, where the ratio $\dfrac{\sigma_f}{\sigma_{CN}}$ is plotted. We assume similar error of 15\% in the fission branching ratios plotted in the lower panel.

\begin{figure}
\begin{adjustwidth}{-0.6in}{-0.6in}
\includegraphics[width=17cm,height=14cm]{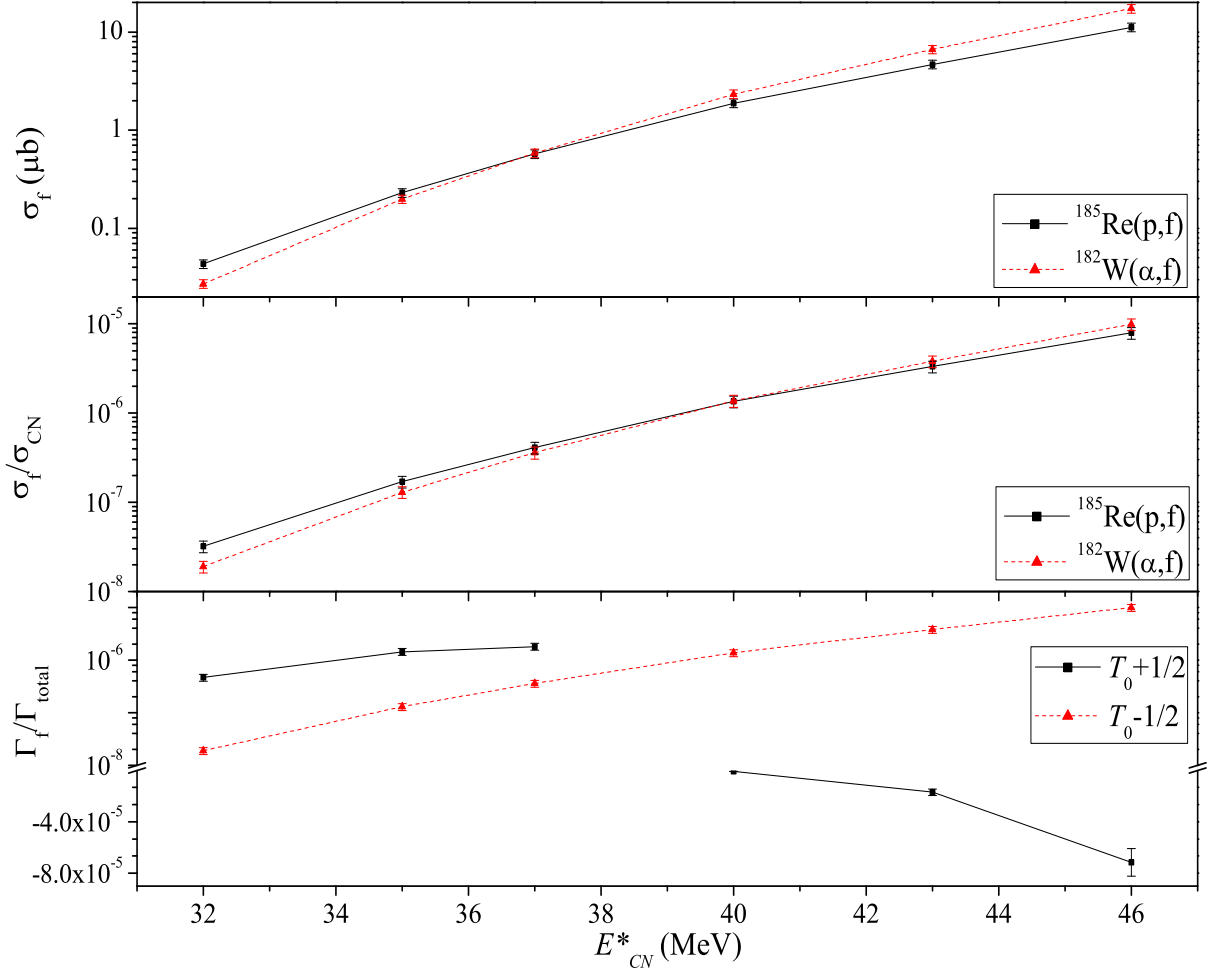}
\caption{\label{fig:186os}(Color online) Same as in Fig.~\ref{fig:210po}, but for a pair of $^{185}$Re($p$, f) and $^{182}$W($\alpha$, f) reactions, both leading to the same compound nucleus $^{186}$Os.}
\end{adjustwidth}
\end{figure}

It is most interesting to see that there is a crossing of the fission cross-sections near $E^{*}_{CN} \approx$ 29 MeV. This crossing leads to negative values of fission branching ratios for $T_0+1/2$ states of CN, a rather unphysical feature. We note that the Eq. (4) used to calculate the fission branching ratios leads to the negative sign. This is a signature of the onset of the region where non-compound processes start making significant contributions to the fission cross-section. The estimates for  $\sigma_{p}$ and $\sigma_{\alpha}$ by using PACE4 may not be good in this region. However, the conclusion that a marked change occurs due to transition from CN to non-CN dominance, remains valid.

Our next data set is a combination of $^{185}$Re($p$, f) and $^{182}$W($\alpha$, f) reactions, both leading to the CN $^{186}$Os. We extract the experimental data of fission cross-sections, $\sigma_{p,f}$ and $\sigma_{\alpha,f}$ from Ignatyuk $\it{et}$ $\it{al.}$~\cite{ignatyuk} as shown in the upper panel of Fig.~\ref{fig:186os}. In this case, the alpha induced fission cross-section begins to dominate the proton induced fission cross-section around $E^{*}_{CN}$= 37 MeV. The ratio of $\dfrac{\sigma_f}{\sigma_{CN}}$ is shown in the middle panel of Fig.~\ref{fig:186os}, and again a crossing occurs near $E^{*}_{CN}$= 40 MeV, as also seen in the previous example. Below this energy, the fission branching ratio for $T_0+1/2$ is approximately ten times greater than that for $T_0-1/2$ state which further supports the idea that isospin is an important factor in fission reactions proceeding through compound nuclear process. 

\begin{figure}
\begin{adjustwidth}{-0.6in}{-0.6in}
\includegraphics[width=17cm,height=14cm]{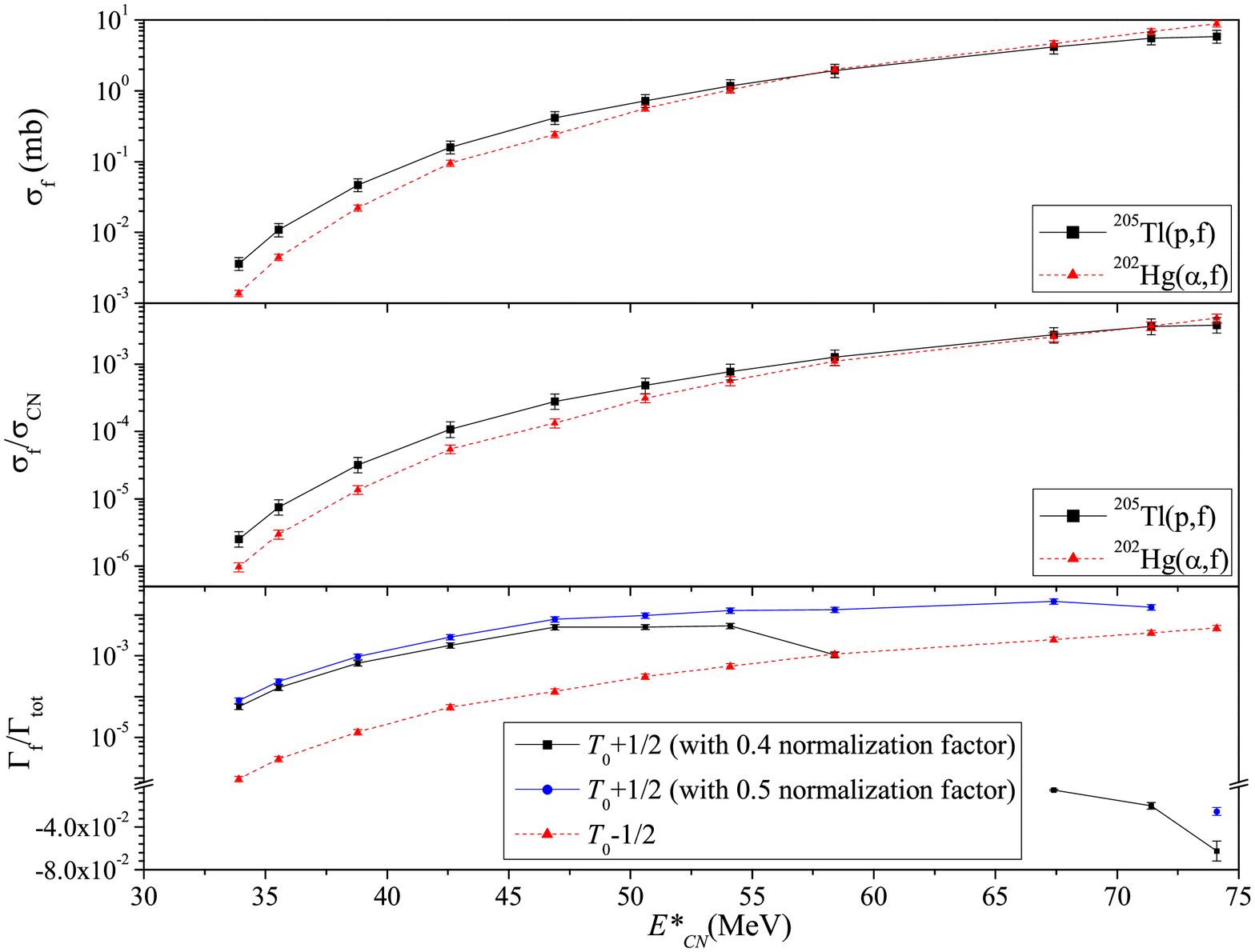}
\caption{\label{fig:206pb}(Color online) Same as in Fig.~\ref{fig:210po}, but for a pair of reactions $^{205}$Tl($p$, f) and $^{202}$Hg($\alpha$, f), both leading to the same compound nucleus $^{206}$Pb. The experimental data for $\sigma_{p,f}$ is plotted with a normalization factor varying between 0.4 and 0.5 (absorbed in the size of the square symbol) and an additional error of 10\% shown by error bars, in the upper panel. The ratio $\dfrac{\sigma_{p,f}} {\sigma_{p}}$ is likewise plotted in the middle panel, but the error bars here reflect a 15\% error.}
\end{adjustwidth}
\end{figure}

Now, we consider the third combination of reactions, namely, $^{205}$Tl($p$, f) and $^{202}$Hg($\alpha$, f), where the CN $^{206}$Pb is formed. The experimental data of fission cross-sections for proton and alpha induced fission are taken from Ignatyuk $\it{et}$ $\it{al.}$~\cite{ignatyuk} and Moretto $\it{et}$ $\it{al.}$~\cite{moretto}, respectively. As pointed out in section 3, we normalize the data of $^{205}$Tl($p$, f) from Ignatyuk $\it{et}$ $\it{al.}$~\cite{ignatyuk} by using factors 0.4 (the lower limit) and 0.5 (the upper limit). We plot these values in the upper panel of Fig. 4 as square symbols and use the size of the square to represent the spread due to normalization uncertainty. Additional error bars are superimposed on these squares, which carry an error of 10\%. These values for $^{205}$Tl($p$, f) are then compared with the fission cross-section from $^{202}$Hg($\alpha$, f). Between $E^{*}_{CN}$= 50-60 MeV, $\sigma_{\alpha,f}$ becomes greater than $\sigma_{p,f}$ and the ratio $\dfrac{\sigma_f}{\sigma_{CN}}$ for alpha induced fission starts to dominate the proton induced fission. Below this excitation energy, there is a difference of the order of one hundred between the fission branching ratios from two isospin states of CN. At higher energies, non-compound processes start making a significant contribution and a gradual transition is seen in the figure.

It may be pointed out that all the three data sets exhibit identical trends and behavior. They lead to the same conclusion that the CN formed in these reactions carries a memory of the isospin state, which in turn influences the fission decay width and the fission cross-section.

\begin{figure}
\begin{adjustwidth}{-0.6in}{-0.6in}
\includegraphics[width=17cm,height=14cm]{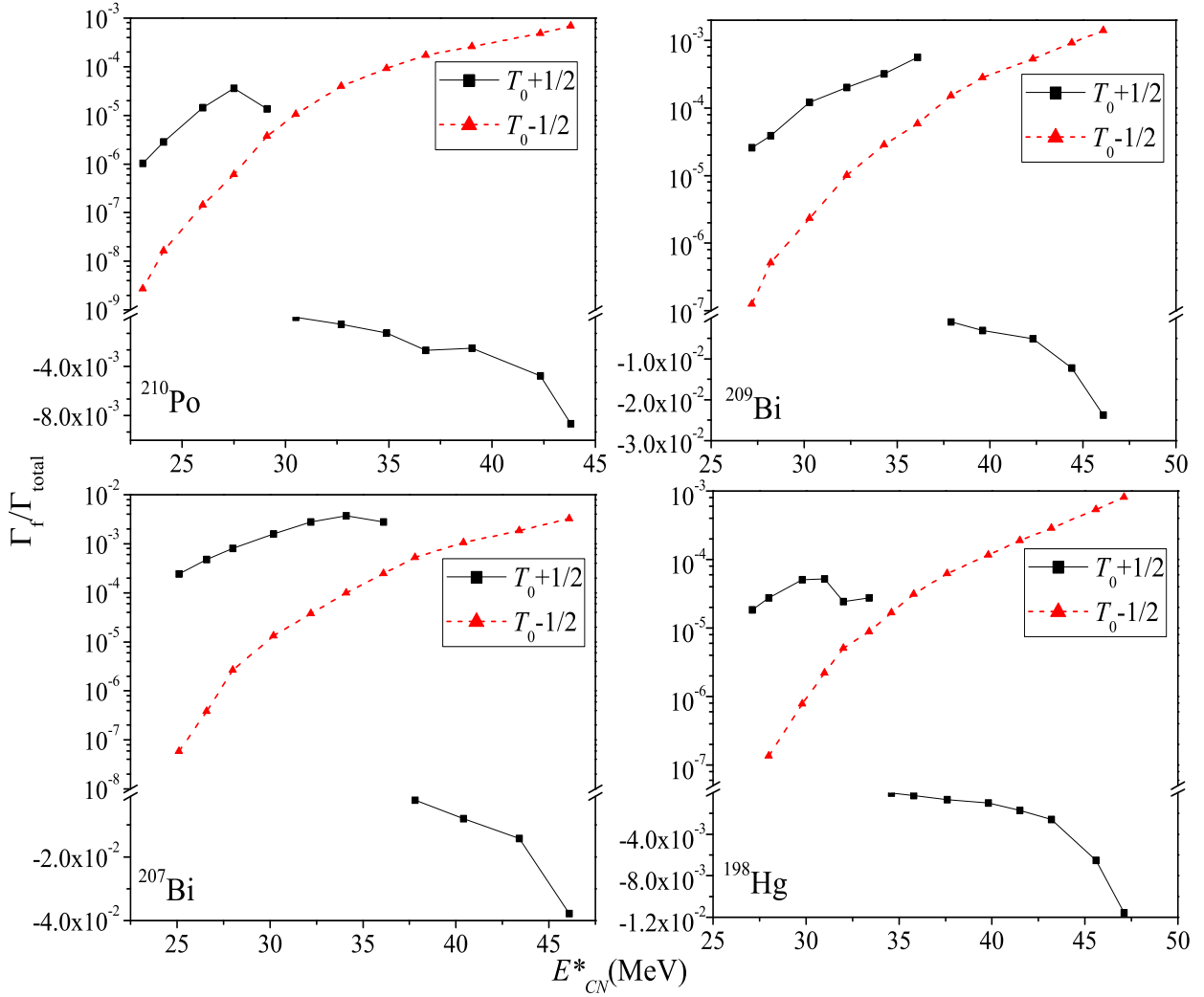}
\caption{\label{fig:igna1977_merge}(Color online) Empirical fission branching ratios for the two different isospin states obtained from the data on fissility for four CN, $^{210}$Po, $^{209}$Bi, $^{207}$Bi and $^{198}$Hg. Experimental data are taken from Zhukova $\it{et}$ $\it{al.}$~\cite{ignatyuk1977b}. We use two scales in all the panels, linear scale before break and log scale after break.}
\end{adjustwidth}
\end{figure}

\textbf{Empirical approach 2 -} In this approach, we perform calculations for the four cases of CN, viz., $^{210}$Po, $^{209}$Bi, $^{207}$Bi and $^{198}$Hg, using the data from Zhukova $\it{et}$ $\it{al.}$~\cite{ignatyuk1977b}. In these cases, we do not require the PACE4 calculations since we have the experimental data for the fissility parameter which are plotted in Fig.~\ref{fig:igna1977}. We calculate the fission branching ratios for the $T_0+1/2$ and $T_0-1/2$ states of CN using Eqs. 3, 4 and 5. We plot the results in Fig.~\ref{fig:igna1977_merge}. We note that $^{210}$Po is the common case in the two approaches. From both the Fig.~\ref{fig:210po} and \ref{fig:igna1977_merge}, we can see that a crossing occurs near $E^{*}_{CN}=29-30$ MeV for the CN $^{210}$Po. The fission branching ratios begin to show an unnatural behavior at $E^{*}_{CN} \geq 30$ MeV. The other three cases also show a large difference between the fission branching ratios from the two isospin states of the same CN. Also, the fission branching ratios become negative near $E^{*}_{CN} \approx 36-38$ MeV for $^{209}$Bi, $E^{*}_{CN} \approx 36-38$ MeV for $^{207}$Bi and $E^{*}_{CN} \approx 33-35$ MeV for $^{198}$Hg. All these calculations confirm the isospin dependence in fission involving heavy $N>Z$ nuclei.

We would like to stress that no entrance channel effect is expected in the light ion fission reactions being considered here. The earlier works on entrance channel effects, for example the detailed work of Saxena $\it{et}$ $\it{al.}$~\cite{saxena}, use the Businaro-Gallone mass asymmetry $\alpha_{BG}$ to check the possibility of entrance channel effects. If the values of the mass asymmetry $\alpha$ for a given pair of reactions leading to the same CN lie on either side of the $\alpha_{BG}$ value, an entrance channel effect is expected to occur. In the pair of reactions that we are considering, $\alpha$ values of both the reactions are all quite large ($>$0.95) compared to $\alpha_{BG}$ =0.7 to 0.8 approx. This supports the non-existence of any entrance channel dependence in the fission cross-sections being considered in our paper.

We would like to emphasize the twin role that large isospin values play. Firstly, the value of the fission branching ratio for the isospin $T_0+1/2$ gets amplified by a factor of ($2T_0+1$) as in Eq. (4). The decay widths, which control the branching ratios, are the physically relevant quantities containing the nuclear structure information and the difference in the two decay widths leads to the different fission cross-sections for the two isospin states. Secondly, the isospin mixing factor of $1/(T+1)$ ensures that isospin remains an approximately good quantum number in heavy nuclei, even though the level densities may be high at the excitation energies being considered. These two factors ensure that the effect will be observable only in $N>Z$ heavy nuclei or, neutron-rich systems where $T$ is large. This also explains why only a few examples are available where this effect may be confirmed. With the availability of rare isotopic beams, it may become possible to test this idea in additional cases.
\section{Conclusions}
\label{sec 4}
In this paper, we present empirical evidence of isospin dependence in fission, which is based on the idea proposed by Yadrovsky~\cite{yadrovsky}, where he calculated the fission decay widths for $T_0+1/2$ and $T_0-1/2$ states of CN for a combination of two reactions $^{209}$Bi($p$, f) and $^{206}$Pb($\alpha$, f) leading to the same CN $^{210}$Po. We also calculate the fission branching ratios for the same set of reactions but with an improved experimental data set and two more combination of reactions, namely, $^{185}$Re($p$, f) and $^{182}$W($\alpha$, f) leading to $^{186}$Os and $^{205}$Tl($p$, f) and $^{202}$Hg($\alpha$, f) leading to $^{206}$Pb. A large difference between the fission branching ratios from the two isospin states clearly suggests that the memory of isospin persists in the CN fission of heavy mass nuclei. 

In the second approach, we use the data of fissility to calculate the fission branching ratios for two isospin states in four cases of CN, namely $^{210}$Po, $^{209}$Bi, $^{207}$Bi and $^{198}$Hg from Zhukova $\it{et}$ $\it{al.}$~\cite{ignatyuk1977b}. The large difference between the fission branching ratios from $T_0+1/2$ and $T_0-1/2$ states of CN for all the four cases suggests an important and effective role of isospin in such systems. We have not used the PACE4 estimates in arriving at this conclusion for these four cases. The evidence for the four cases is, therefore, based purely on the experimental data. At this stage, we would like to leave this conclusion as a distinct possibility requiring further experimentation and analysis.

The large isospin values are observed to play a crucial role in heavy nuclei. On the one hand, large $T$ ensures purity of isospin by a factor of $1/(T+1)$, and on the other hand, it amplifies the effect by a factor of $2T+1$, making it observable in $N>Z$ heavy nuclei. Such an effect should also become prominent in neutron-rich nuclei, which will become accessible by rare isotope beams.

At higher energies, where the non-compound processes start making a significant contribution, the picture changes gradually and the fission branching ratio for $T_0+1/2$ state becomes negative. The assumption of CN formation is not strongly valid at these energies. This crossing possibly provides a new signature of the transition, where the non-compound processes begin to dominate the compound processes. 

New experimental efforts should be made to obtain more precise data for the cases discussed in this paper and test these ideas. In addition to this, there are many incomplete data sets in the early works of Igantyuk $\it{et}$ $\it{al.}$ which may be completed~\cite{ignatyuk1977b,ignatyuk,ignatyuk1976,ignatyuk1977a}. Few examples of pairs of reactions having incomplete data in Ignatyuk $\it{et}$ $\it{al.}$ ~\cite{ignatyuk1977b,ignatyuk} are: $^{196}$Pt($p$, f) and $^{193}$Ir($\alpha$, f) forming $^{197}$Au, $^{188}$Os($p$, f) and $^{185}$Re($\alpha$, f) forming $^{189}$Ir and $^{190}$Os($p$, f) and $^{187}$Re($\alpha$,f) forming $^{191}$Ir.  More specifically, we propose that new experiments may be carried out for the following pair of reactions, viz., $^{206}$Pb($d$, f) and $^{207}$Pb($p$, f), $^{200}$Hg($p$, f) and $^{197}$Au($\alpha$, f), $^{235}$U($p$, f) and $^{234}$U($d$, f).
\section*{Acknowledgement}
We thank T. K. Ghosh (VECC, Kolkata), R. Palit (TIFR, Mumbai) for valuable comments and suggestions. We also thank Arun Jain (BARC, Mumbai) for many helpful discussions. Support from the Ministry of Human Resource Development (Government of India) to SG in the form of a fellowship is gratefully acknowledged.





\end{document}